# Comparing Spark vs MPI/OpenMP
# On Word Count MapReduce

*Junhao Li*
*Department of Physics, Cornell University*

Spark provides an in-memory implementation of MapReduce that is widely used in the big data industry. MPI/OpenMP is a popular framework for high performance parallel computing. This paper presents a high performance MapReduce design in MPI/OpenMP and uses that to compare with Spark on the classic word count MapReduce task. My result shows that the MPI/OpenMP MapReduce is an order of magnitude faster than Apache Spark.

## Overview

MapReduce is a popular paradigm for parallel computing. Many big data processing routines can be transformed into a series of MapReduce tasks and get efficiently executed on highly optimized MapReduce infrastructures.

Spark is one of the MapReduce implementations that features an in-memory implementation, which gives it significantly higher performance and sets it apart from others, making it one of the most popular packages for data analysis.

On the other hand, MPI/OpenMP has long been the standard for high performance computing. However, there is no mature MapReduce implementation with MPI/OpenMP.

In this paper, I describe a high performance implementation of MapReduce in MPI/OpenMP. Preliminary result shows that my design achieves an order of magnitude speedup compared to Apache Spark on the classic work frequency count task.

This design is mainly based a distributed hash table from Arrow, our recently developed quantum chemistry program. The usage of the distributed hash table along with other innovations boosted the performance of that program by several orders of magnitude.

## MPI/OpenMP MapReduce Design

At a high-level, my design has three key data types: DistRange, DistHashMap, and ConcurrentHashMap.

DistRange can be constructed by providing the start, end, and step size. DistRange provides a distributed map method that will map the numbers in the range to the available threads. The mapper function takes the number and pushes zero or more entries to an instance of DistHashMap.

DistHashMap is a simplified DHT (distributed hash table) that only ensures eventual consistency for associative inserts / updates. For a cluster of n nodes, a DistHashMap consists of, on each node, a main ConcurrentHashMap to store all the data entries belong to the current node, and (n - 1) additional ConcurrentHashMaps to store the data belong to other nodes but inserted / updated by the current node and pending synchronization.

ConcurrentHashMap is hash map that supports efficient and thread safe insertions / updates by an arbitrary number of threads on a single node. It consists of a data portion and a thread cache portion. The data portion consists of several linear probing hash maps, called segments. Each segment is responsible for storing a certain hash range in the entire hash space. When a thread wants to update a segment, it has to lock the segment first. In the case that a segment is already locked by another thread, the data will be flushed to a thread-local linear-probing hash map in the thread cache portion, so that no thread will ever get blocked. The usage of linear probing hash maps can give high performance in a shared memory setting because it incurs less memory allocation and bulk memory access than chained hash tables, which is the default in many STL implementations (C++ standard library). The cache will be synchronized to the main data portion either periodically or after the map phase ends.

The inter-node synchronization will also be performed either periodically or after the map phase ends. After the map phase ends, all the nodes start to shuffle the data to the correct node and upon receiving the new data, the main ConcurrentHashMap inserts the new data into itself in parallel.

## Word Count Result

Word count is a classic MapReduce task where the input is an English text consisting of words separated by spaces and the output is the number of occurrences of each word. The map function takes a portion of the text and emits (word, 1) pairs to a distributed map. The reduce function is simply the summation (by key).

Word count is given as an example on Spark's website. The source code is as follows:

```
val input = sc.textFile("...")
val output = input.flatMap(line => line.split(" "))
                  .map(word => (word, 1))
                  .reduceByKey(_ + _)
```

My MPI/OpenMP MapReduce implementation is available at [https://github.com/jl2922/fgpl/tree/wordcount2](https://github.com/jl2922/fgpl/tree/wordcount2). The entrance point of the word count is at src/test/dist_range_test.cc. The high level interface is as follows:

```
DistRange<int> range(0, lines.size());
DistHashMap<std::string, int> target;
const auto& mapper = [&](const int i, const auto& emit) {
  std::stringstream ss(lines[i]);
  std::string word;
  while (std::getline(ss, word, ' ')) {
    emit(word, 1);
  }
};
range.mapreduce<std::string, int, std::hash<std::string>>(
    mapper, Reducer<int>::sum, target);
```

I run both Spark's word count and my MPI/OpenMP implementation on exactly the same hardware on AWS (Amazon Web Service). The Spark cluster is set up with the default settings of AWS EMR (Elastic MapReduce). At the time of the test, Amazon offers EMR 5.20.0 which comes with Spark 2.4.0. My MPI/OpenMP implementation is compiled with G++ 7.2 and MPICH 3.2. For both implementations, I use AWS EC2 (Elastic Computing Cloud) r5.xlarge instances (4 vCPU, 32 GB RAM).

The input text is from the Bible and Shakespeare's works, repeated about 200 times to make it roughly 2 GB in size.

Here are the results (converted to words per second):

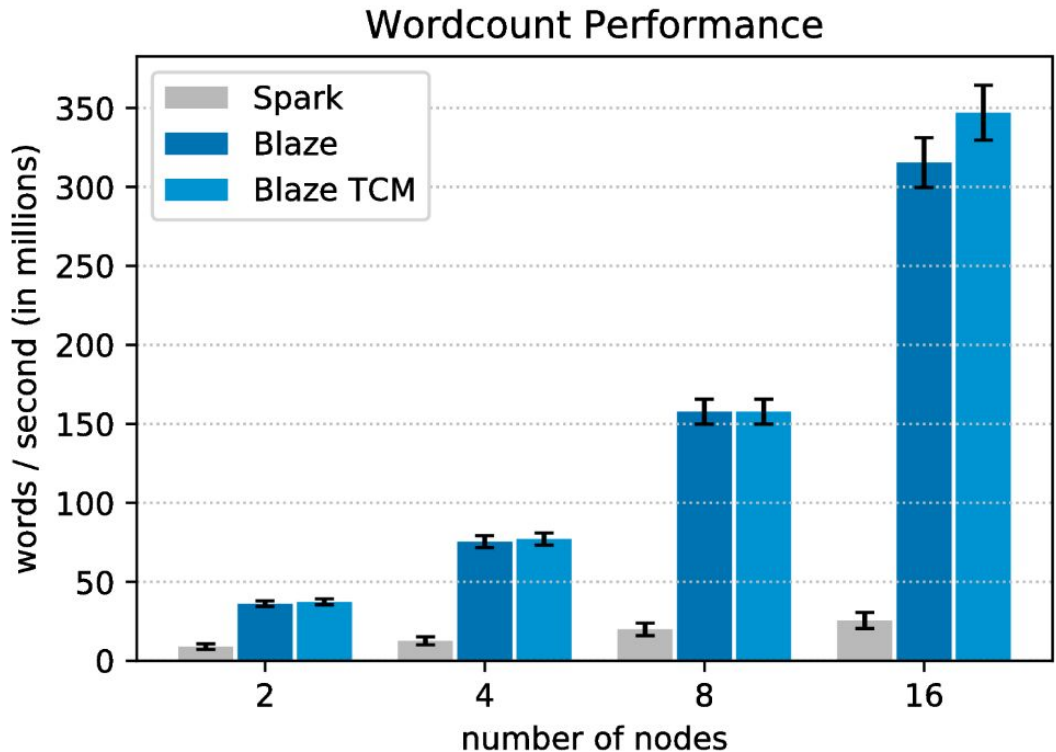

Here Blaze is the published code name of the library that includes my implementation. TCM means the code is linked with the TCMalloc memory allocator.

We can see that my MPI/OpenMP design is an order of magnitude faster than Spark/Scala.

There are several possible reasons for this:

- MPI/OpenMP uses C++ and runs natively while Spark/Scala runs through a virtual machine.

- MPI/OpenMP is not designed for fault tolerance, so my design does not consider that while Spark does. Fault tolerance incurs additional overhead.
- My design performs local reduce during the map phase before shuffling the (key, value) pairs so that the network traffic is significantly reduced.

## Conclusion

MPI/OpenMP has extremely high performance and can outperform Spark/Scala on its most proficient task by an order of magnitude.

For cases where we do not care about fault tolerance and only care about performance or costs, MPI/OpenMP may be a much better choice than Spark. This also includes most offline data analysis use cases, such as business intelligence, where the running time of a task is usually much less than a million core hours. The MTBF (mean time between failures) on modern hardware is on the scale of one million core hours, so the failure rate for these tasks should be extremely low even without fault tolerance, and in the case of a rare failure, we can simply run the task multiple times and as long as it succeeds before the fourth try, we are still likely to get the results faster than using Spark.

## Acknowledgements

This work is supported by the U.S. National Science Foundation (NSF) grant ACI-1534965 and the Air Force Office of Scientific Research (AFOSR) grant FA9550-18-1-0095. We also thank professor Cyrus Umrigar for the helpful suggestions for the paper.